\documentclass[useAMS,usenatbib]{mn2e}
\usepackage{graphicx}
\title[Propylene Oxide and Glycine in Sagittarius B2 and Orion]
{A Search for Propylene Oxide and Glycine in Sagittarius B2 (LMH) and 
Orion}
\author[M. R. Cunningham et al.]{M. R. Cunningham$^{1}$,
\thanks{E-mail: Maria.Cunningham@unsw.edu.au (MRC)} 
P. A. Jones$^{1}$, P. D. Godfrey$^{2}$, D. M. Cragg$^{2}$,  I. Bains$^{1,3}$, \newauthor{M.G. Burton$^{1}$, P. Calisse$^{1}$, N. H. M. Crighton$^{1}$, S. J. Curran$^{1}$, T. M. Davis$^{1}$,} \newauthor{J. T. Dempsey$^{1}$, B. Fulton$^{1}$, M. G. Hidas$^{1}$, T. Hill$^{1}$, L. Kedziora-Chudczer$^{1}$, }  \newauthor{V. Minier$^{1}$, M. B. Pracy$^{1}$, C. Purcell$^{1}$, J. Shobbrook$^{1}$ and T. Travouillon$^{1}$}  \\
$^{1}$School of Physics, University of New South Wales, NSW 2052, Australia \\
$^{2}$School of Chemistry, Monash University, Clayton, Victoria 3800, 
Australia \\
$^{3}$Swinburne University of Technology, PO Box 218, Hawthorn, Victoria 3122, Australia
}

\begin{document}

\date{Accepted . Received ; in original form }

\pagerange{\pageref{firstpage}--\pageref{lastpage}} \pubyear{2007}

\maketitle

\label{firstpage}

\begin{abstract}

We have used the Mopra Telescope to search for glycine and the simple chiral molecule propylene oxide in the Sgr B2 (LMH) and Orion KL, in the 3-mm band. We have not detected either species, but have been able to put sensitive upper limits on the abundances of both molecules. The 3--sigma upper limits derived for glycine conformer I are 3.7 $\times 10^{14}$ cm$^{-2}$ in both Orion-KL and Sgr B2 (LMH), comparable to the reported detections of  conformer I  by Kuan et al. However, as our values are 3-sigma upper limits rather than  detections we conclude that this weighs against confirming the detection of Kuan et al. We find upper limits for the glycine II column density of 7.7 $\times 10^{12}$ cm$^{-2}$ in both Orion-KL and Sgr B2 (LMH), in agreement with the results of Combes et al. The results presented here show that glycine conformer II is not present in the extended gas at the levels detected by Kuan et al. for conformer I. Our ATCA results (Jones et al.) have ruled out the detection of glycine (both conformers I and II) in the compact hot core of the LMH at the levels reported, so we conclude that it is unlikely that  Kuan et al. have detected glycine in either Sgr B2 or  Orion--KL. We find upper limits for propylene oxide abundance of 3.0 $\times 10^{14}$ cm$^{-2}$ in Orion-KL and 6.7 $\times 10^{14}$ cm$^{-2}$ in Sgr B2 (LMH). We have detected fourteen features in Sgr B2 and four features in Orion-KL which have not previously been reported in the ISM, but have not be able to plausibly assign these transitions to any carrier.

\end{abstract}

\begin{keywords}
ISM:molecules - radio lines:ISM - ISM:individual:Sgr B2 - ISM:individual:Orion
\end{keywords}

\section[]{Introduction}

%Citation Definitions and Commands %%%%%%%%%%%%%% 
\defcitealias{paper1}{Paper I} 	  	 
%\citetalias{jon90} 	-> 	Paper I
%\citepalias{jon90} 	-> 	(Paper I)
%%%%%%%%%%%%%%% 

There is now a substantial body of work to suggest that life on Earth arose from an initial reservoir of simpler organic material that formed in the presolar nebula rather than on the early Earth (see e.g Holtom et al. 2005; Bailey et al. 1998; Hunt-Cunningham \& Jones 2004). These arguments centre around 
\begin{enumerate}
\item primitive mechanisms of self-replication involving only RNA,  which indicate the need for a reservoir of homochiral (having like chirality\footnote{Chiral molecules are those where asymmetry in carbon atom placement within a molecule leads to two distinct forms (mirror images or enantiomers), often referred to as left-handed or right-handed. An excess of about 10 \% in one enantiomer over the other (L over D in the case of amino acids on the Earth) is needed for RNA template reproduction to work.}) organic molecules on the early Earth before living organisms could evolve (Joyce et al. 1984), and 
\item the lack of any known abiotic mechanisms operating on the surface of the Earth to generate the required enantiomeric excess  \citep{bonner99,bonner91}. 
\end{enumerate}

However, an extraterrestrial origin for this reservoir is not universally accepted, and mechanisms which could amplify an enantiomeric excess on planetary surfaces, such as catalytic processes on the surfaces of minerals, have also been suggested \citep[see][and references therein]{ehrenfreund}.  The detection of chiral molecules in the interstellar medium (ISM) would lend support to an extraterrestrial origin for the initial reservoir of prebiotic molecules. Despite a possible extraterrestrial origin for homochiral molecules, none of the molecules so far detected in the interstellar medium (ISM) is chiral. 

\begin{figure} 
\includegraphics{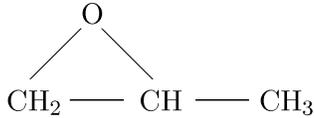}
\caption{The chemical structure of the simple chiral molecule propylene oxide ($\rm{c-CH_{3}C_{2}H_{3}O}$)}
\label{fig1}
\end{figure}

One relatively simple chiral molecule that may be present in detectable quantities in the ISM is propylene oxide (c-C$_3$H$_6$O) (Fig \ref{fig1}). Its non-chiral lower homologue, ethylene oxide (c-C$_2$H$_4$O), has been detected in the ISM (Dickens et al. 1997). In addition, propylene (C$_3$H$_6$), although not yet detected in the ISM, is considered important as an intermediary in the production of ISM polycyclic aromatic hydrocarbons (Kaiser et al. 2001). Propylene oxide is known to form in the gas phase from the oxidation of propylene \citep[see][and references therein]{zuwei01}, although the likelihood of this mechanism in interstellar contexts has not yet been studied. The known interstellar homologue of propylene oxide, ethylene oxide, is isomeric with two other known interstellar species: acetaldehyde (CH$_3$CHO) and vinyl alcohol (CH$_2$=CHOH). The likelihood of interstellar propylene oxide is supported by the interstellar detection \citep{hollis+04} of its isomer propanal (CH$_3$CH$_2$CHO).

%figure 1%%%%%%%%%%%%%%%
%%%%%%%%%%%%%%% 

Another group of molecules of particular interest for the origin of life are the amino acids, which are generally chiral and which are likely to have been present in the presolar nebula. Glycine, although not a chiral molecule, is the simplest amino acid utilized in biological processes on Earth, and is found in carbonaceous chondrite meteorites, which are thought to be relatively unprocessed remnants of the presolar nebula (see e.g. Wirick et al. 2006). However, despite numerous searches covering more than a quarter of a century (see e.g. Kuan et al. 2004) glycine has not yet been incontrovertibly detected in the ISM (Snyder et al. 2005). This is in spite of the fact that we now know of more than 130 molecular species in the ISM, including molecules of similar complexity, such as acetic acid, which shares some structures with glycine. The reported detection by \citet{k+2003} was disputed by \citet{snyder05}, who concluded that if glycine was present in the compact hot cores of Sgr B2 (N-LMH), Orion KL and W51 e1/e2 then it could only be in the form of the higher energy conformer II.  In Jones et al. (2006, hereafter \citetalias{paper1}), we showed, using observations from the Australia Telescope Compact Array (ATCA), that glycine conformer II could not be present at the levels reported by \citet{k+2003} for Sgr B2 (N-LMH) if it was confined to small scales on the order of the LMH continuum source ($<$ 5 arcsec).

In this paper we report on a search for glycine (conformer II) and propylene oxide in more extended gas using the Mopra telescope\footnote{The Mopra Radio Telescope, located in Coonabarabran NSW, is run by the Australia Telescope National Facility on behalf of the Commonwealth Science and Industry Research Organisation (CSIRO), Australia.}. We have also serendipitously observed several transition frequencies of glycine conformer I. The search was initially prompted by the new availability of laboratory measured (rather than calculated) transition frequencies for both molecules in the 86-115 GHz range \citepalias{paper1}, allowing the search to be undertaken with high frequency resolution. The search has involved a multi-pronged approach, using both ATCA and the Mopra telescope, as it was not clear whether the molecules were most likely to be found in small diameter hot cores, in which case an interferometer is the best strategy, or whether they were more likely to be extended, in which case a single dish telescope is the preferred instrument. \citetalias{paper1} gives a detailed justification of the search strategy.  Most previous searches for glycine have been for glycine I, so we have concentrated on glycine II. This conformer has relatively stronger transitions than conformer I, owing to its larger dipole moment, although it is higher in energy \citepalias{paper1}. Under laboratory conditions glycine II is more readily detected in the higher temperature gas, and the expected line intensities of the two conformers equalise at 285 K \citep{hollis}. Under interstellar conditions the energy barrier between conformers I and II will inhibit relaxation from one to the other, so the relative abundance of the two conformers will depend on the chemical and thermal history of the gas.

\section[]{Mopra Observations and Data Reduction}
\label{obs}
Single-dish observations in the 3-mm band were made with the 22-m Mopra 
telescope of the Australia Telescope National Facility, during 2002 and 2003\footnote{The data reported here were taken as part of shared operations of the telescope between the University of New South Wales (UNSW) and CSIRO}.
There were two pointing positions used in each of Sgr B2 and Orion. The positions (J2000) 
were a) Sgr B2 (LMH) 17 47 19.8, -28 22 17.0 ($\alpha=0'',\delta=0''$) b) nearby 17 47 21.28, -28 22 17.0 ($\alpha=+26'',\delta=0''$), c) Orion-KL 05 35 14.5, -05 22 29.6 ($\alpha=0'',\delta=0''$) and d) nearby 05 35 15.8, -05 22 29 ($\alpha=+19'',\delta=0''$). The log of observations is given in Table 1. 

%table 1

\begin{table*}
\caption{Log of Mopra observations, frequency bands and lines searched.
The band around 96413 MHz (*) was covered with 2 different tunings. The frequencies quoted are from new laboratory measurements made at the School of Chemistry, Monash University. ID specifies the
molecule, propylene oxide (PO) or glycine (G), and conformer, I or II (glycine only), with this transition frequency. Strength
is a measure of the intensity predicted at 10 K and 50 K (10 K if only
one number is given), relative to a value of 100 for the strongest
line of that species, at the stated temperature, in the 80-115 GHz range. }
\begin{tabular}{rlcrll}
\hline
Band  &  Dates        & Source & Frequency  & ID  &  Strength \\
      &               &        &   (MHz)      &     &  10/50 K  \\  
\hline
86262  & 2002 Oct      & Sgr B2 & 86262.23  & G II $12_{6,7} - 11_{6,6}$ & 27/51 \\
86847  & 2002 Jul      & Sgr B2 & 86960.81  & G II $14_{1,14} - 13_{1,13}$ & 74/89 \\
       &               &        & 86967.06  & G II $14_{0,14} - 13_{0,13}$ & 74/89 \\
88348  & 2002 Sep      & Orion, Sgr B2 & 88348.31   & PO E $4_{2,2} - 3_{1,3}$   & 48 \\
       &               &    & 88348.62   & PO A $4_{2,2} - 3_{1,3}$   & 48 \\
91337  & 2002 Aug - Nov & Orion, Sgr B2 & 91337.94 & PO   $7_{1,7} - 6_{0,6}$   & 69 \\
92881  & 2003 Aug, Sep & Orion & 92881.427  & G II $13_{9,5} - 12_{9,4}$ & 4/29 \\
       &               &       & 92881.429  & G II $13_{9,4} - 12_{9,3}$ & 4/29 \\
94045  & 2002 Sep      & Orion, Sgr B2 & 94045.74   & PO $8_{0,8} - 7_{1,7}$ & 53 \\
	&		&		& 94043.84	& G I $14_{3,12} - 13_{3,11}$ & 52/72 \\
94623  & 2002 Sep      & Orion, Sgr B2 & 94623.93   & PO E $5_{2,4} - 4_{1,3}$ & 59 \\
       &               &    & 94624.21   & PO A $5_{2,4} - 4_{1,3}$ & 59 \\
96413* & 2002 Jul - Sep & Orion, Sgr B2 & 96414.25  & G I $7_{4,3} - 6_{3,4}$ & 64 \\
		&	 &  & 96421.14  & PO $3_{3,1} - 2_{2,0}$ & 94 \\
	&                &    & 96422.14  & PO $3_{3,0} - 2_{2,0}$ & 11 \\
100332 & 2003 Aug      & Orion & 100332.264 & G II $14_{8,7} - 13_{8,6}$ & 7/44 \\ 
       &               &       & 100332.509 & G II $14_{8,6} - 13_{8,5}$ & 7/44 \\ 
102188 & 2002 Sep, Oct & Orion, Sgr B2 & 102188.83  & PO $8_{1,8} - 7_{0,7}$ & 69 \\
103094 & 2002 Sep      & Orion, Sgr B2 & 103094.22  & PO $5_{2,3} - 4_{1,4}$ & 51 \\
105436 & 2002 Sep, Oct & Orion, Sgr B2   & 105436.55  & PO $6_{2,5} - 5_{1,4}$ & 61 \\
107013 & 2002 Oct      & Orion & 107006.19  & PO $9_{0,9} - 8_{1,8}$ & 51 \\
108983 & 2002 Sep, Nov & Orion, Sgr B2 & 108983.31  & PO $4_{3,2} - 3_{2,1}$ & 100 \\
109160 & 2002 Sep, Nov & Orion, Sgr B2 & 109160.72  & PO $4_{3,1} - 3_{2,2}$ & 100 \\
113153 & 2002 Sep      & Orion, Sgr B2 & 113153.11  & PO $9_{1,9} - 8_{0,8}$ & 60 \\
       &               &    & 113155.91 & G I $17_{2,15} - 16_{2,14}$ & 22/94 \\
% 114972 & 2002 Sep      &    & 114981.774 & G II $9_{3,6} - 8_{1,7}$ & 6 \\
%        &               &    & 114984.746 & G II $6_{6,1} - 5_{5,0}$ & 7 \\
%        &               &    & 114984.783 & G II $6_{6,0} - 5_{5,1}$ & 7 \\
%        &               &    & 114999.895 & G II $16_{8,9} - 15_{8,8}$ & 4/52 \\
%        &               &    & 115002.023 & G II $16_{8,8} - 15_{8,7}$ & 4/52 \\
\hline
\end{tabular}
\label{tab1}
\end{table*}

% table 2

\begin{table*}
\caption{Parameters of Mopra spectra.
The positions in column 4 are (J2000), with offsets given in arcsec:
 a) Sgr B2 (LMH) 17 47 19.8, -28 22 17.0 ($\alpha=0'',\delta=0''$) b) nearby 17 47 21.28, -28 22 17.0 ($\alpha=+26'',\delta=0''$), c) Orion-KL 05 35 14.5, -05 22 29.6 ($\alpha=0'',\delta=0''$) and d) nearby 05 35 15.8, -05 22 29 ($\alpha=+19'',\delta=0''$). The RMS noise temperatures are quoted in terms of  $T_A$ rather than $T_A^*$, that is to say they have not been corrected for the Mopra beam efficiency (see text). 
}
\begin{tabular}{rrllrrc}
\hline
Band  & Tuning Freq. & Source & Pos. & RMS & Integr. & mean \\
      & (MHz)        &        &     & (mK) & time (s) & T$_{sys}$ (K) \\  
\hline
 86262 &  86262.23 & Sgr B2   & a  & 13 &  2519 & 179 \\
 86847 &  86847.00 & Sgr B2   & a  & 13 & 25733 & 435 \\
 88348 &  88348.00 & Orion    & cd & 12 &  5366 & 240 \\
       &  88348.00 & Sgr B2   & b  & 19 &  2628 & 267 \\
 91337 &  91337.94 & Orion    & cd & 12 & 10731 & 272 \\
       &  91337.94 & Sgr B2   & b  & 14 &  6461 & 250 \\
 92881 &  92881.43 & Orion    & cd &  4 & 57597 & 282 \\
 94045 &  94045.74 & Orion    & d  & 18 &  2409 & 231 \\
       &  94045.74 & Sgr B2   & b  & 17 &  5307 & 319 \\
 94623 &  94623.93 & Orion    & d  & 19 &  3066 & 264 \\
       &  94623.93 & Sgr B2   & b  &  8 & 13359 & 240 \\
 96413 &  96413.00 & Orion    & cd & 19 &  7825 & 373 \\
       &  96413.00 & Sgr B2   & a  & 18 &  5854 & 288 \\
 96440 &  96440.00 & Orion    & d  & 12 &  5694 & 251 \\
       &  96440.00 & Sgr B2   & b  & 13 &  8103 & 238 \\
100332 & 100332.87 & Orion    & d  & 13 &  4380 & 234 \\ 
102188 & 102188.83 & Orion    & d  & 11 & 10950 & 295 \\
       & 102188.83 & Sgr B2   & b  & 19 &  3942 & 334 \\
103094 & 103094.22 & Orion    & d  & 22 &  2409 & 281 \\
       & 103094.22 & Sgr B2   & b  & 15 &  2847 & 197 \\
105436 & 105436.55 & Orion    & cd & 14 &  3614 & 238 \\
       & 105436.55 & Sgr B2   & b  & 17 &  3504 & 269 \\
107013 & 107013.85 & Orion    & d  & 21 &  3504 & 349 \\
108983 & 108983.31 & Orion    & cd &  9 & 10961 & 253 \\
       & 108983.31 & Sgr B2   & b  & 20 &  7328 & 368 \\
109160 & 109160.72 & Orion    & cd & 17 &  9437 & 322 \\
       & 109160.72 & Sgr B2   & b  & 52 &  1832 & 517 \\
113153 & 113153.11 & Orion    & d  & 15 & 14656 & 368 \\
       & 113153.11 & Orion    & d  & 37 &  1832 & 450 \\
       & 113153.11 & Sgr B2   & b  & * & 17404 & 436 \\
\hline
\multicolumn{7}{c}{* No meaningful RMS could be calculated for this spectrum due to contamination from the $^{13}$CO line in the lower sideband.} \\
\end{tabular}
\label{tab2}
\end{table*}

The Mopra telescope had an SIS receiver for the 3-mm band that required tuning
with a phase-locked oscillator and adjustment of the mixers for good sensitivity
and side-band rejection. To cover the 16 bands of Table 1, we mostly tuned 
to the frequency of the propylene oxide or glycine line, as shown in Table 2,
but in a few cases we used a slightly different frequency to get a more stable 
tuning. For example, for the band 86847 we used the well-used and tested 
SiO line tuning. We used two different tunings around 96421 MHz, that is 
96413 and 96440, giving a total of 17 different tunings in 
Table 2. 

The observations covered a frequency range of 64 MHz, with 1024 channels, and 
two orthogonal linear polarisations A and B. Pointing observations were made
every hour or so using SiO masers, by retuning polarisation B to the SiO 
frequency. The system temperature was measured by a chopper wheel system at
the start of each long integration in the schedule file.
Typically each scan was for 110 s, with on-source and off-source reference
scans alternating in the pattern off-on-on-off.

The data were initially reduced with the ATNF SPC package, designed to handle 
the ATNF format of RPFITS data file.
Individual source scans had the off-source reference subtracted and were
averaged over the period of the individual RPFITS file of up to an hour.
We used the DFM\footnote{http://www.phys.unsw.edu.au/astro/mopra/software.php} graphical interface to SPC (written by Cormac Purcell) to do this processing, and 
used a perl script, moprafix.pl\footnote{http://www.atnf.csiro.au/people/cphillip/software.html}, to correct
the RPFITS headers for a known small frequency error due to rounding off of the
rest frequency information. The spectrum from each file was inspected and
good spectra were combined together for a given tuning, velocity setting and 
source (Sgr B2 or Orion) to produce the 30 spectra listed in Table \ref{tab2}. The 
combination was done with weighting depending on the system temperature,
and after small pixel shifts to put the spectra on the same LSR velocity scale.

Because of problems with stability in tuning to these
unusual frequencies, not all spectra were good, presumably because of a loss 
of lock in the local oscillator. Polarisations A and B were combined in about 
half the cases, but for many files it appears that polarisation B was not good,
probably because of the tuning back and forth to the SiO maser frequency for pointing.

The Orion spectra showed little or no change between the two pointing centres and so were averaged together, in some cases, to increase sensitivity, as they were within a beamwidth (see Table \ref{tab2}) .
For the tuning 96413 in Sgr B2 the observations were 
made with two different frequencies, due to two different LSR velocities being assumed for the 
source, and so appear as two different spectra.

The spectra had a linear baseline fitted and subtracted, and were then
Hanning smoothed.
The parameters of the spectra are summarised in Table 2, with integration time,
weighted mean system temperature and RMS noise level in line-free regions
of the spectra. The RMS noise temperatures quoted in Table \ref{tab2} are in terms of  $T_A$ rather than $T_A^*$, (where $T_A^*= T_A/\eta_{MB}$), that is to say they have not been corrected for the Mopra beam efficiency, which varies from 0.49 at 78 GHz to 0.42 at 115 GHz \citep{ladd}. For consistency with Figures \ref{fig2} and \ref{fig3}, the values for peak and integrated intensity in Tables \ref{tab3}, \ref{tab4} and \ref{tab5} are also quoted in units of $T_A$ rather than $T_A^*$. The RMS noise level decreases with integration time as expected with time $t$ as $t^{-1/2}$, indicating that we are not affected by systematic baselevel ripple effects, but the system temperatures are in practice rather high due to many observations at large zenith distance and poor weather.

\section[]{Results}

\begin{figure*}
\setlength{\unitlength}{1cm}
\caption{Spectra of Orion. The offset shown is in arcsec from Orion-KL $\alpha$ = 05 35 14.5, $\delta$ = -05 22 29.6. The units of brightness temperature are $T_A$ rather than $T_A^*$ and have not been corrected for the Mopra beam efficiency. The expected positions of transitions of propylene oxide (PO), Glycine conformers I (GI) and II (GII) are marked, with the label shown in brackets to distinguish these from detected features.}

\includegraphics[width = 19 cm]{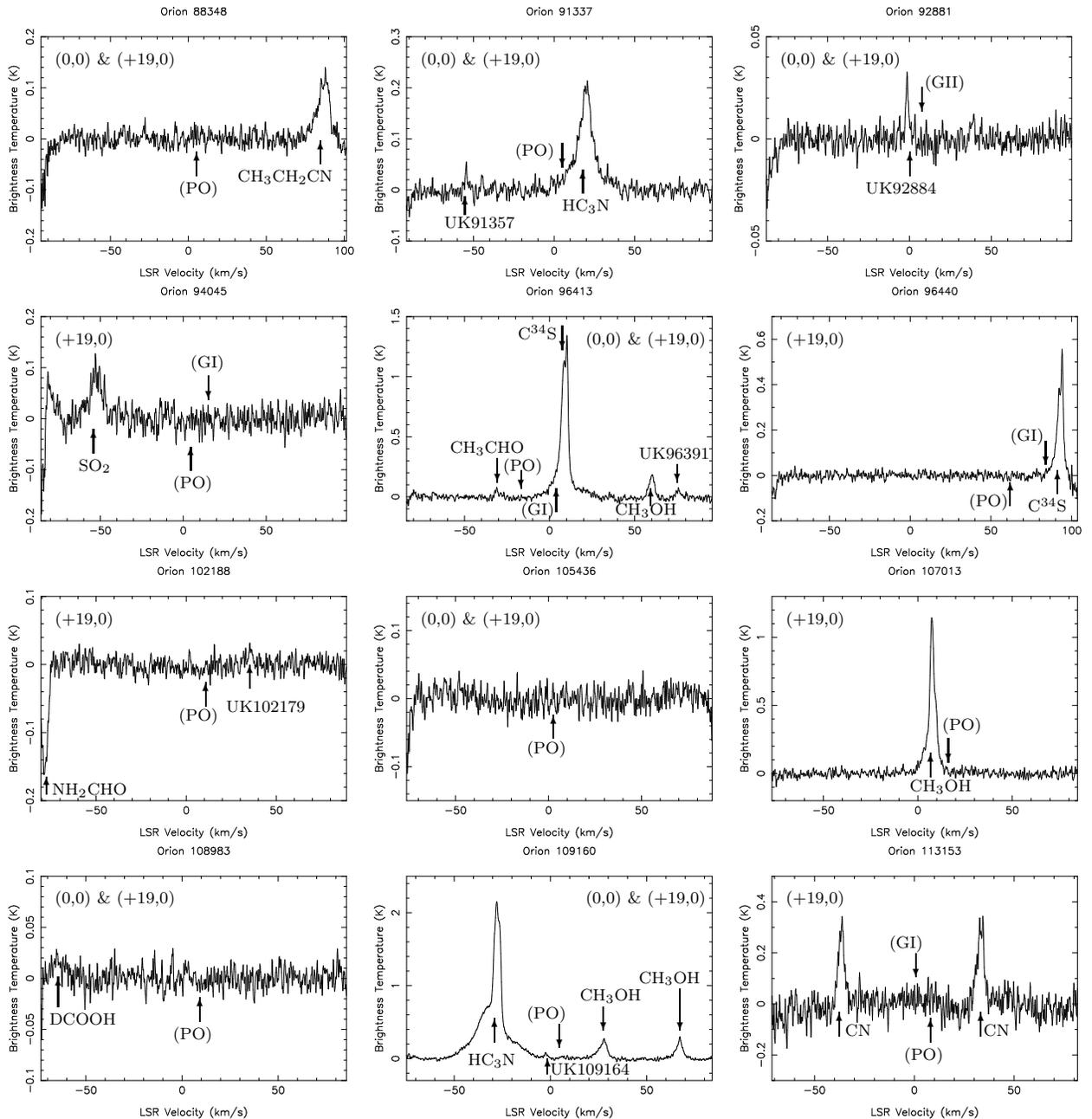}

\label{fig2}
\end{figure*}

\begin{figure*}
\setlength{\unitlength}{1cm}
\caption{Spectra of Sgr B2 (LMH). The offset shown is in arcsec from Sgr B2 (LMH) $\alpha$ = 17 47 19.8, $\delta$ = -28 22 17.0. The units of brightness temperature are $T_A$ rather than $T_A^*$ and have not been corrected for the Mopra beam efficiency. The expected positions of transitions of propylene oxide (PO), Glycine conformers I (GI) and II (GII) are marked, with the label shown in brackets to distinguish these from detected features.}

\includegraphics[width = 19 cm]{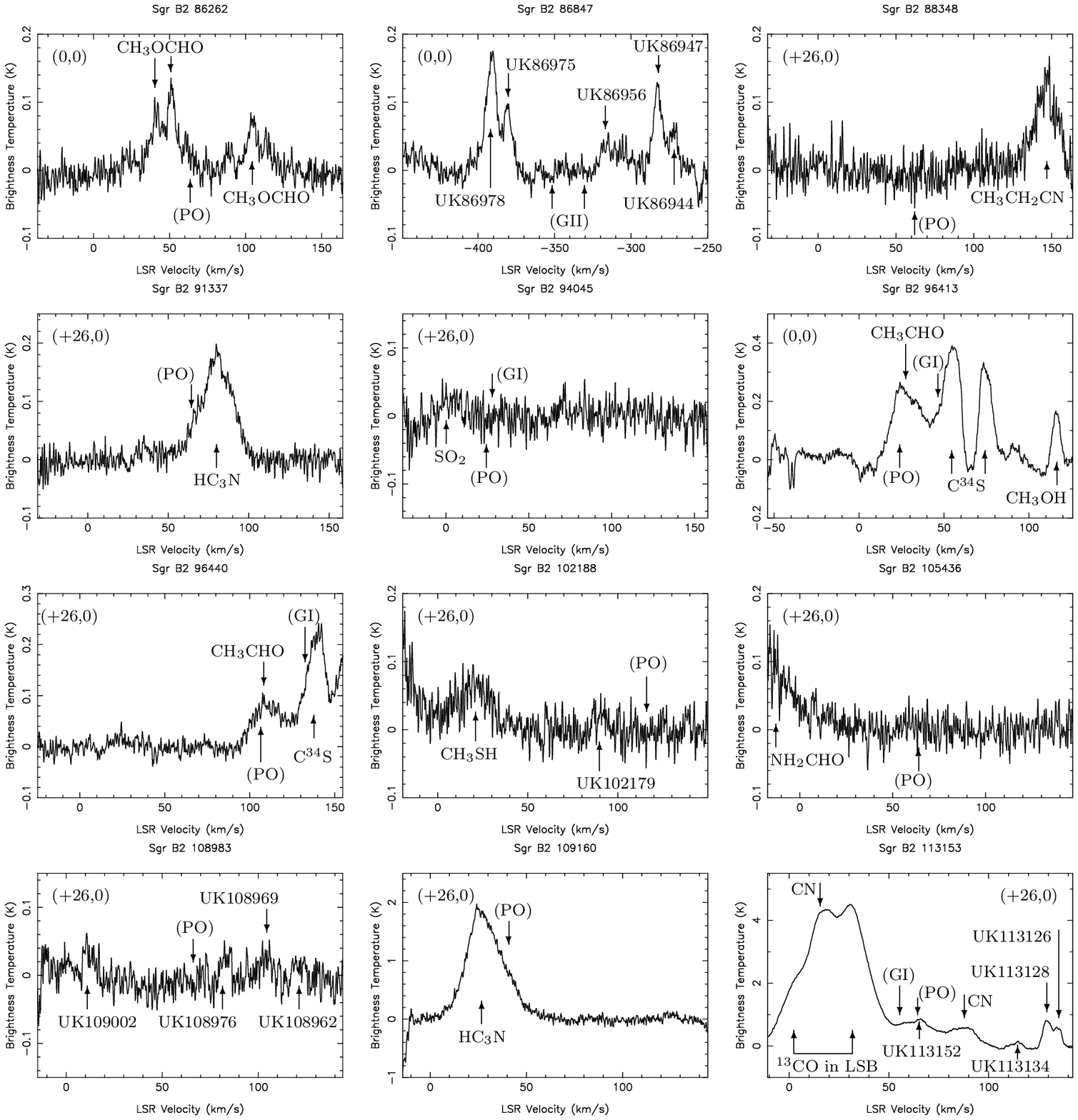}

\label{fig3}
\end{figure*}

The spectra are plotted in Figures \ref{fig2} and \ref{fig3}, with the exception of five 
spectra that just showed noise, namely tunings 94623, 100332 and 103094 
for Orion and tunings 94623 and 103094 for Sgr B2. The spectra are plotted
dropping 100 channels from each end, where the bandpass correction can
lead to spurious features, but some edge effects may still remain.

Most of the spectra show some line features, which were fitted with Gaussians,
as summarised in Tables 3 and 4. The LSR velocity scales of the spectra are calculated
for the rest frequency of the tuning in Table 2. Most of the lines detected
here are listed in the NIST on-line database of Lovas 
(http://physics.nist.gov/PhysRefData/), and we give the line identification
in column 8 of Tables 3 and 4. There are 18 features in Tables 3 and 4 which do not have a corresponding entry in Lovas. These lines have been labeled for convenience in discussion with the prefix UK followed by the frequency in MHz.

%%%%% FIGURE 3 %%%%%%%%%%

% table 3

\begin{table*}
\caption{\textbf{Orion-KL:} Emission and absorption lines detected above the 3 $\sigma$ level (See fig \ref{fig2}.) The peak and integrated brightness temperatures are quoted in $T_A$ rather than $T_A^*$ and have not been corrected for the Mopra beam efficiency.
}
\begin{tabular}{rrrrrcll}
\hline
Band &  Peak & Centre & FWHM & Integrated & Rest freq & ID comment \\
     &         (mK) & (km s$^{-1}$) & (km s$^{-1}$) & (K km s$^{-1}$) & (MHz)       &          \\
\hline

88348 &  113 &   86.6 &  8.9 & 1.07 & 88324.8 & CH$_3$CH$_2$CN 88323.754 \\
%	&	&	&	&	&	&	\\
91337 &   51 &  -54.5 &  1.2 & 0.07 & 91357.0 & UK91357 \\ 
      &         176 &   19.6 & 11.8 & 2.21 & 91334.4 & HC$_3$N 91333.308 \\
%&	&	&	&	&	&	\\
92881 &   29 &   -1.4 &  2.0 & 0.06 & 92884.3 & UK92884, ($^{29}$SiC$_2$ at 92882.18) \\
%&	&	&	&	&	&	\\
94045 &   87 &  -52.4 &  8.8 & 0.82 & 94064.7 & SO$_2$ 94064.686 \\
%&	&	&	&	&	&	\\
96413 &   50 &  -30.9 &  3.0 & 0.16 & 96425.5 & CH$_3$CHO 96425.62 \\
      &         1200 &   9.0 &  5.0 & 6.40 & 96412.7 & C$^{34}$S 96412.961 \\
      &         178 &   60.1 &  3.6 & 0.68 & 96396.2 & CH$_3$OH 96396.055 \\
      &          47 &   75.8 &  2.9 & 0.15 & 96391.2 & UK96391 \\ 
%&	&	&	&	&	&	\\
96440 &  442 &   93.2 &  4.3 & 2.02 & 96412.6 & C$^{34}$S 96412.961 \\
%&	&	&	&	&	&	\\
102188 &  -156 & -78.3 & 3.6 & -0.60 & 102218.2 & NH$_2$CHO abs 102217.571 \\
       &          24 &  35.6 &  2.6 & 0.07 & 102179.4 & UK102179 \\
%&	&	&	&	&	&	\\
107013 &  984 &   7.6 &  4.1 & 4.30 & 107014.0 & CH$_3$OH 107013.77 \\
%&	&	&	&	&	&	\\
108983 &   17 & -65.0 &  6.8 & 0.12 & 109009.8 & DCOOH 109008.67 \\
%&	&	&	&	&	&	\\
109160 &   647 & -31.0 & 19.1 & 13.2 & 109174.9 & HC$_3$N broad 109173.638 \\
       &         1590 & -27.6 &  3.5 & 5.91 & 109173.7 & HC$_3$N narrow 109173.638 \\
       &           58 &  -2.5 &  3.3 & 0.20 & 109164.5 & UK109165 \\ 
       &          228 &  27.3 &  5.4 & 1.31 & 109153.7 & CH$_3$OH 109153.21 \\
       &          241 &  67.0 &  4.2 & 1.08 & 109139.2 & CH$_3$OH 109137.57 \\
%&	&	&	&	&	&	\\
%113153 &   204 &  31.4 &  6.3 & 1.37 & 113144.3 & CN 113144.192 \\
%       &         8580 & 134.5 &  4.1 & 37.5 & 113105.4 & 110201.35 $^{13}$CO in LSB \\ 
%&	&	&	&	&	&	\\
113153 &   301 & -36.5 &  3.5 & 1.12 & 113169.9 & CN 113170.528 \\
       &          299 &  33.2 &  4.2 & 1.34 & 113143.6 & CN 113144.192 \\
\hline
\end{tabular}
\label{tab3}
\end{table*}

% table 4

\begin{table*}
\caption{\textbf{Sgr B2:} Emission and absorption lines detected above the 3 $\sigma$ level (See fig \ref{fig3}). We have assumed a cloud velocity of 64 km s$^{-1}$ to convert the observed frequency to rest frequency. The peak and integrated brightness temperatures are quoted in $T_A$ rather than $T_A^*$ and have not been corrected for the Mopra beam efficiency.}
\begin{tabular}{rrrrrcrl}
\hline
Band &  Peak & Centre & FWHM & Integrated & Rest freq &  & ID comment \\
     &         (mK) & (km s$^{-1}$) & (km s$^{-1}$) & (K km s$^{-1}$) & (MHz)       &          \\
\hline
86262 &   75 &   40.5 &  8.4 & 0.67 & 86269.0 &  & CH$_3$OCHO 86268.659 \\
      &         103 &   51.5 &  9.0 & 0.99 & 86265.8 &  & CH$_3$OCHO 86265.826 \\
      &          52 &  106.3 & 18.0 & 1.00 & 86250.1 &  & CH$_3$OCHO 86250.576 \\
%&	&	&	&	&	&	\\
86847 &         165 & -391.2 &  8.6 & 1.51 & 86978.9 &  & UK86979 \\
      &          90 & -380.3 &  6.0 & 0.58 & 86975.7 &  & UK86976 \\
      &          29 & -313.7 & 23.9 & 0.74 & 86956.4 &  & UK86956 \\ 
      &         118 & -282.6 &  7.1 & 0.89 & 86947.4 &  & UK86947 \\ 
      &          59 & -272.2 &  6.6 & 0.41 & 86944.4 &  & UK86944 \\
%&	&	&	&	&	&	\\
88348 &  120 &  146.2 & 17.2 & 2.20 & 88323.8 &  & CH$_3$CH$_2$CN  88323.754 \\
%&	&	&	&	&	&	\\
91337 &  169 &   80.2 & 23.5 & 4.23 & 91333.0 &  & HC$_3$N 91333.308 \\
%&	&	&	&	&	&	\\
94045 &   25 &    3.8 & 10.7 & 0.28 & 94064.6 &  & SO$_2$ 94064.686 \\
96413 &  252 &   29.5 & 13.1 & 3.52 & 96424.1 &  & CH$_3$CHO 96425.62 \\
      &         363 &   53.5 & 14.3 & 5.53 & 96416.4 &  & C$^{34}$S - self abs 96412.961 \\
      &         340 &   74.7 &  7.5 & 2.72 & 96409.6 &  & C$^{34}$S - self abs 96412.961 \\
      &         172 &  116.2 &  4.4 & 0.81 & 96396.2 &  & CH$_3$OH  96396.055 \\
%&	&	&	&	&	&	\\
96440 &   83 &  111.7 & 21.8 & 1.93 & 96424.7 &  & CH$_3$CHO 96425.62 \\
      &         219 &  139.3 & 14.3 & 3.34 & 96415.8 &  & C$^{34}$S - self abs 96412.961 \\
%&	&	&	&	&	&	\\
102188 &   58 &  18.5 & 28.7 & 1.77 & 102204.3 &  & CH$_3$SH 102202.49 \\
       &          24 &  90.4 &  7.3 & 0.19 & 102179.8 &  & UK102180 \\ 
%&	&	&	&	&	&	\\
105436 &   89 & -10.4 & 13.4 & 1.27 & 105462.7 &  & NH$_2$CHO at edge 105464.216 \\
%&	&	&	&	&	&	\\
108983 &          49 &  11.2 &  7.0 & 0.37 & 109002.5 &  & UK109003 \\ 
       &          43 &  83.4 &  4.9 & 0.22 & 108976.3 &  & UK108976 \\ 
       &          35 & 103.7 & 10.7 & 0.40 & 108968.9 &  & UK108969 \\ 
       &          24 & 121.3 &  9.8 & 0.25 & 108962.5 &  & UK108963 \\
%&	&	&	&	&	&	\\ 
109160 &  1780 &  28.1 & 22.5 & 42.6 & 109173.8 &  & HC$_3$N 109173.638 \\
%&	&	&	&	&	&	\\
113153 &  1340 &   2.3 & 15.5 & 22.2 & 113176.4 &  & 110201.35 $^{13}$CO in LSB \\ 
       &         1140 &  14.6 &  8.3 & 10.1 & 113171.8 &  & CN 113170.528 \\
       &         4400 &  27.0 & 30.8 &  144 & 113167.1 &  & 110201.35 $^{13}$CO in LSB \\ 
       &          722 &  66.2 & 14.7 & 11.3 & 113152.3 &  & UK113152 \\
       &          603 &  86.3 & 16.0 & 10.3 & 113144.7 &  & CN 113144.192 \\
       &          131 & 114.6 &  3.5 & 0.49 & 113134.0 &  & UK113134, (N$^{34}$S at 113136.20) \\
       &          841 & 129.4 &  4.5 & 4.03 & 113128.4 &  & UK113128 \\
       &          591 & 134.7 &  4.3 & 2.71 & 113126.4 &  & UK113126 \\
\hline
\end{tabular}
\label{tab4}
\end{table*}

We also checked our spectra by eye with those
from the surveys of Sgr B2 and Orion in Turner (1989). We have assumed 
systemic LSR velocities of 8 km s$^{-1}$ for Orion and 64 km s$^{-1}$ for Sgr B2
in calculating the rest frequencies of our fitted line features in column 7 of 
Tables 3 and 4. However, given the range 
of velocities within both clouds, this will lead to some uncertainty. 
For 17 strong lines in Orion with good line identifications, the mean 
difference of the rest frequencies here and in the NIST database is
0.3 MHz with standard deviation 0.7 MHz, corresponding to mean velocity
difference of 0.9 km s$^{-1}$ with standard deviation 2.1 km s$^{-1}$. Similarly
for 10 lines in Sgr B2 the mean difference is 0.0 MHz with standard deviation 
0.3 MHz, or difference 0.0 km s$^{-1}$ with standard deviation 0.9 km s$^{-1}$.

Due to the sensitivity of these observations, we detect lines in most of the 
spectra (Tables 3 and 4), but none of these lines can be identified with the 
propylene oxide or glycine lines searched for (Table 1). 

Due to the possible problems with tuning stability, we have done a careful
check of the Mopra spectra for known lines. Out of 30 portions of spectrum observed
we have detected lines in 23, and of these 20 include features with good identifications
of one or more lines from the NIST database. Of the 3 portions of spectrum with lines detected here, but with no lines within the band listed in 
the NIST database, the 86847 Sgr B2 spectrum shows new weak lines that agree
with our observations with the Australia Telescope Compact Array \citepalias{paper1}.  The higher intensity of these lines (UK86978, UK86975 and UK86956) in the ATCA observations compared to those from the Mopra Telescope is consistent with the carriers being confined to the LMH. UK86975 is likely to be from the same carrier as UK86978, as the $\sim$3 MHz  ($\sim$9 km s$^{-1}$) separation is consistent with that of the two velocity components of 63.5 and 73.1 km s$^{-1}$ found towards the Sgr B2 LMH. The other two portions of spectrum without lines in the NIST database (Sgr B2 108983 and Orion 92881) have weak lines, which plausibly have not been detected previously.
There are also unidentified lines in the 20 spectra with NIST
identifications. Of the six spectra with no lines detected above the noise,
one is the Orion spectrum at 105436 MHz with a strong line at its edge
(NH$_2$CHO) in the corresponding Sgr B2 spectrum, so we are confident of this 
tuning. The other five spectra (tunings 94623 and 103094 for Orion and Sgr B2 
and 100332 for Orion (not shown in Figures \ref{fig2} and \ref{fig3}) are questionable, as the NIST database lists CH$_3$OCHO lines at 94632.735 and 103114.88
MHz and HC$_3$N at 100322.4 MHz. The expected strength of these lines from
other observations (eg. Turner 1991) and rotation diagram fits to other lines
of these molecules is (subject to some uncertainty) comparable to, or above the
3--sigma limits. Therefore, we have not used these 5 spectra in calculating 
the limits for the propylene oxide and glycine II lines (Table \ref{tab5}). We also
note that tuning 113153 in Sgr B2 is contaminated by strong
 $^{13}$CO 110201.353 emission from the lower sideband, 3.0 GHz below
the upper sideband, in a tuning with poor sideband rejection.

\section[]{Limits for propylene oxide and glycine II}
\label{limits}

The upper limits for transitions of propylene oxide and glycine II are listed 
in Table \ref{tab5}.

Using standard assumptions of LTE radiative transfer (eg. Rohlfs \& Wilson 2004)
the column density $N_u$ of molecules in the upper level of the transition
is related to the line intensity by
\[ N_u = (8 \pi \nu^2 k/h c^3 A_{ul}) \int T_B dv \]
where $A_{ul}$ is the Einstein coefficient, and $\int T_B dv$ is the 
integral over velocity of the brightness temperature $T_B$ of the emission line.
The Einstein coefficients $A_{ul}$ for glycine
were obtained from the Pickett JPL 
on-line database \citep{p+98} from the tabulated log of intensity $I(T_o)$ at 
reference temperature
$T_o = 300$~K. $\int T_B dv$ for each transition was calculated assuming a line width of 4.5 km s$^{-1}$ for Orion KL and 8.0 km s$^{-1}$ for Sgr B2. The upper limit to $T_{MB}$ was calculated using three times the RMS value for the appropriate tuning from Table \ref{tab2} corrected  for the Mopra main beam efficiency (see section \ref{obs}), with an extra correction for the number of independent velocity pixels compared to the expected line width. The uncertainty in the peak flux of a Gaussian fit to data is smaller than the RMS of individual pixels, because several individual pixels combine together. The values in Table \ref{tab2} for the RMS are for the Hanning smoothed spectra, which have an independent velocity pixel width of 0.75 km s$^{-1}$, giving 6 independent points across the assumed line width in Orion--KL and 10.7 points in Sgr B2. The final value used for the upper limit to $\int T dv$ was thus corrected by $1/\sqrt{n}$, where $n$ is the number of independent pixels.

The total column density $N$ of the molecule is given by
\[ N = (N_u/g_u) Q_T \exp(E_u/kT_{ex}) \]
where $Q_T$ is the partition function at excitation temperature $T_{ex}$, $E_u$
is the energy of the upper level and $g_u$ is the statistical weight of the 
upper level. Values for $g_u$, the upper state energy $E_u$, frequency $\nu$ and
Q as a function of temperature for glycine were also obtained 
from the JPL database. 

%table 5

\begin{table*}
\caption{Limits for lines searched, and comments on lines
where limits were not obtained. We label with `conf.' the transitions which were
confused, and the confusing lines. We list flux density and brightness 
temperature limits, molecular line parameters $A_{ul}$, $g_{u}$, $E_{u}/k$, 
and column density limits
for the upper level of the transition $N_{u}$ and the total $N$. The peak and integrated brightness temperatures are quoted in $T_A$ rather than $T_A^*$ and have not been corrected for the Mopra beam efficiency, but these corrections were applied before calculating the upper limits.}
\begin{tabular}{lcrcccrrcc}
\hline
 Source & ID & Frequency & 3 $\sigma$ limit & 3 $\sigma$ limit & $A_{ul}$ &
$g_u$ & $E_u/$k & $N_u$ & $N$ \\
        &    & (MHz)~~~   & (mK) & (K km s$^{-1}$) & ($s^{-1}$) & 
  & (K)~ & (cm$^{-2}$) & (cm$^{-2}$) \\
\hline

Orion & G I & 94043.84 & 54 & 0.259 & $3.69 \times 10^{-6}$ & 87 & 37.3 &
$2.6 \times 10^{12}$ & $5.4 \times 10^{14}$ \\
Orion & G I & 113155.91 & 45 & 0.216 & $6.56 \times 10^{-6}$ & 105 & 52.5 &
$1.9 \times 10^{12}$ & $3.7 \times 10^{14}$\\
      &      &           &    &       &   &  & \\
Sgr B2 & G I & 94043.84 & 51 & 0.435 & $3.69 \times 10^{-6}$ & 87 & 37.3 & 
$4.3 \times 10^{12}$ &  $3.7 \times 10^{14}$\\
      &      &           &    &       &   &  & \\
Orion & G II & 92881.427 & 12 & 0.058 & $6.80 \times 10^{-5}$ & 81 & 1063.2 &
$3.0 \times 10^{10}$ & $7.7 \times 10^{12}$ \\
%Orion & G II & 92881.429 & 12 & 0.058 & $6.80 \times 10^{-5}$ & 81 & 1063.2 &
%$1.4 \times 10^{10}$ & \\
      &      &           &    &       &   &  & \\
Sgr B2 & G II & 86262.231 & 39 & 0.332 & $7.81 \times 10^{-5}$ & 75 & 1044.7 &
$1.3 \times 10^{11}$ & $1.3 \times 10^{13}$ \\
Sgr B2 & G II & 86960.812 & 39 & 0.332 & $1.06 \times 10^{-4}$ & 87 & 1038.7 &
$9.7 \times 10^{10}$ & $7.7 \times 10^{12}$ \\
Sgr B2 & G II & 86967.057 & 39 & 0.332 & $1.06 \times 10^{-4}$ & 87 & 1038.7 &
$9.7 \times 10^{10}$ & $7.7 \times 10^{12}$ \\
       &     &          &    &     \\
Orion & PO & 88348.31 & 36 & 0.173 & $ 3.83 \times 10^{-6}$ &  9 &  8.3 &
$ 1.4 \times 10^{12}$ & $ 1.2 \times 10^{15}$ \\
Orion & PO & 88348.62 & 36 & 0.173 & $ 3.83 \times 10^{-6}$ &  9 &  8.3 &
$ 1.4 \times 10^{12}$ & $ 1.2 \times 10^{15}$ \\
Orion & PO & 91337.94 & 36 & 0.173 & $ 7.98 \times 10^{-6}$ & 15 & 17.0 &
$ 7.5 \times 10^{11}$ & $ 4.0 \times 10^{14}$ \\
Orion & PO & 94045.74 & 54 & 0.259 & $ 8.71 \times 10^{-6}$ & 17 & 21.5 &
$ 1.1 \times 10^{12}$ & $ 5.3 \times 10^{14}$ \\
Orion & PO & 96421.14 & 36 & 0.173 & $ 1.03 \times 10^{-5}$ &  7 &  8.7 &
$ 6.6 \times 10^{11}$ & $ 7.2 \times 10^{14}$ \\
Orion & PO & 96422.14 & 36 & 0.173 & $ 1.18 \times 10^{-6}$ &  7 &  8.7 &
$ 5.9 \times 10^{12}$ & $ 6.3 \times 10^{15}$ \\
Orion & PO & 102188.83 & 33 & 0.158 & $ 1.17 \times 10^{-5}$ & 17 & 21.7 &
$ 6.3 \times 10^{11}$ & $ 3.0 \times 10^{14}$ \\
Orion & PO & 105436.55 & 42 & 0.201 & $ 7.18 \times 10^{-6}$ & 13 & 15.0 &
$ 1.4 \times 10^{12}$ & $ 8.3 \times 10^{14}$ \\
Orion & PO & 107006.19 & 63 & 0.302 & $ 1.37 \times 10^{-5}$ & 19 & 26.8 &
$ 1.1 \times 10^{12}$ & $ 5.0 \times 10^{14}$ \\
Orion & PO & 108983.31 & 27 & 0.129 & $ 1.20 \times 10^{-5}$ &  9 & 11.1 &
$ 5.8 \times 10^{11}$ & $ 5.0 \times 10^{14}$ \\
Orion & PO & 109160.72 & 51 & 0.244 & $ 1.20 \times 10^{-5}$ &  9 & 11.1 &
$ 1.1 \times 10^{12}$ & $ 9.3 \times 10^{14}$ \\
Orion & PO & 113153.11 & 45 & 0.216 & $ 1.66 \times 10^{-5}$ & 19 & 26.9 &
$ 7.5 \times 10^{11}$ & $ 3.4 \times 10^{14}$ \\
      &     &          &    &       \\
Sgr B2 & PO & 88348.31 & 36 & 0.307 & $ 3.83 \times 10^{-6}$ &  9 &  8.3 &
$ 2.5 \times 10^{12}$ & $ 1.6 \times 10^{15}$ \\
Sgr B2 & PO & 88348.62 & 36 & 0.307 & $ 3.83 \times 10^{-6}$ &  9 &  8.3 &
$ 2.5 \times 10^{12}$ & $ 1.6 \times 10^{15}$ \\
Sgr B2 & PO & 94045.74 & 51 & 0.435 & $ 8.71 \times 10^{-6}$ & 17 & 21.5 &
$ 1.9 \times 10^{12}$ & $ 6.7 \times 10^{14}$ \\
Sgr B2 & PO & 102188.83 & 57 & 0.486 & $ 1.17 \times 10^{-5}$ & 17 & 21.7 &
$ 1.9 \times 10^{12}$ & $ 6.8 \times 10^{14}$ \\
Sgr B2 & PO & 105436.55 & 51 & 0.435 & $ 7.18 \times 10^{-6}$ & 13 & 15.0 &
$ 2.9 \times 10^{12}$ & $ 1.4 \times 10^{15}$ \\
Sgr B2 & PO & 108983.31 & 60 & 0.511 & $ 1.20 \times 10^{-5}$ &  9 & 11.1 &
$ 2.3 \times 10^{12}$ & $ 1.5 \times 10^{15}$ \\
Sgr B2 & PO & 109160.72 & 156 & 1.329 & $ 1.20 \times 10^{-5}$ & 9 & 11.1 &
$ 6.0 \times 10^{12}$ & $ 3.8 \times 10^{15}$ \\
\hline

Orion & G I & 96414.25 &   \multicolumn{2}{c}{conf. C$^{34}$S} \\
	&      &            &  &         \\
Sgr B2 & G I & 96414.25 &   \multicolumn{2}{c}{conf. C$^{34}$S} \\
Sgr B2 & G I & 113155.91 &   \multicolumn{2}{c}{conf. 113152 U} \\
      &      &            &  &       \\
Orion & G II & 100332.264 &   \multicolumn{2}{c}{bad tuning ?} \\
Orion & G II & 100332.509 &   \multicolumn{2}{c}{bad tuning ?} \\
      &      &            &  &         \\
Orion & PO & 94623.93 &  \multicolumn{2}{c}{bad tuning ?} \\
Orion & PO & 94624.21 &  \multicolumn{2}{c}{bad tuning ?} \\
Orion & PO & 103094.22 &  \multicolumn{2}{c}{bad tuning ?} \\
      &    &           &  &              \\
Sgr B2 & PO & 91337.94 &  \multicolumn{2}{c}{conf. HC$_3$N} \\
Sgr B2 & PO & 94623.93 &  \multicolumn{2}{c}{bad tuning ?} \\
Sgr B2 & PO & 94624.21 &  \multicolumn{2}{c}{bad tuning ?} \\
Sgr B2 & PO & 96421.14 &  \multicolumn{2}{c}{conf. CH$_3$CHO} \\
Sgr B2 & PO & 96422.14 &  \multicolumn{2}{c}{conf. CH$_3$CHO} \\
Sgr B2 & PO & 103094.22 &  \multicolumn{2}{c}{bad tuning ?} \\
Sgr B2 & PO & 113153.11 &  \multicolumn{2}{c}{conf. 113152 U} \\
\hline
\end{tabular}
\label{tab5}
\end{table*}

% table 6
\begin{table*}
\caption{Summary of upper limits/ reported detections for glycine conformers I and II in Orion and Sgr B2. The limits given are assuming that the emission was extended with respect to the beam of the telescope used for the observations.
}
\begin{tabular}{llrll}
\hline
Conformer  & Source & Abundance & Telescope & Reference\\
      &        &  (cm$^{-2}$)     &      \\  
\hline
Glycine I & Orion-KL & $<3.7 \times 10^{14}$ & Mopra & This paper \\
Glycine I & Orion-KL & $4.4 \times 10^{14}$ & Kitt Peak 12--m & Kuan et al. (2003) \\
\\
Glycine I & Sgr B2 (LMH)& $<3.7 \times 10^{14}$ & Mopra & This paper \\
Glycine I & Sgr B2 (LMH)& $<1.4 \times 10^{15}$ & ATCA$^1$ & Paper 1 \\
Glycine I & Sgr B2 (LMH)& $4.2 \times 10^{14}$ & Kitt Peak 12--m & Kuan et al. (2003) \\
\\
Glycine II & Orion-KL & $<7.7 \times 10^{12}$ & Mopra & This paper \\
Glycine II & Orion SiO & $<8 \times 10^{12}$ & IRAM 30--m & Combes et al. (1996) \\
\\
Glycine II & Sgr B2 (LMH)& $<7.7 \times 10^{12}$ & Mopra & This paper \\
Glycine II & Sgr B2 (LMH)& $<8.6 \times 10^{13}$ & ATCA$^1$ & Paper 1 \\
Glycine II & Sgr B2 (OH)& $<1 \times 10^{13}$ & IRAM 30--m & Combes et al. (1996) \\
\\
\hline
\multicolumn{5}{c}{Notes: 1. The synthesised beam of the ATCA observations was $17.0 \times 3.4$ arcsec$^{-2}$}\\

\end{tabular}
\label{tab6}
\end{table*}
% end table 6

Since there are several hyperfine components blended together for 
each line, with different Einstein coefficients $A_{ul}$
and statistical weights $g_u$, we quote $N_u$ based on the main level, with 
$A_{ul}$ and $g_u$ for this level. We include all of the hyperfine
components in the calculation for $N$, by considering $\sum A_{ul} g_u$ 
for all the components. We have assumed excitation temperature 
$T_{ex} = 75$ K in Sgr B2 (LMH) and $T_{ex} = 150$ K for Orion-KL, yielding Q(75) = 29377
and Q(150) = 68673. These temperatures were chosen to be similar to those
estimated by Kuan et al. (2003) for glycine I, in Sgr B2 and Orion, to enable 
more direct comparison with that paper and \citetalias{paper1}.

The conformer
II is higher in energy than conformer I by $W_c = 705 $~cm$^{-1}$ (Lovas et al. 1995), with energy 
expressed as $1/\lambda = E/hc$,  and with 
uncertainty 10 \%, or $E/k = 1014 \pm 100$~K. The energy levels of glycine II 
are expressed relative to the lowest energy state of glycine I in the JPL 
database and Table 5, but we have considered the two conformers as separate 
species in the total column density calculations. Hence when correcting for the 
Maxwell-Boltzmann distribution of levels we converted the energies to those
relative to the lowest energy state of glycine II for the $\exp(E_u/kT_{ex})$ 
factor.  Although we primarily concentrated on glycine II, we also serendipitously observed glycine I transitions in three tunings, namely 94045, 96413 and 113153. We were not able to obtain an upper limit for the transition at 96414 MHz in the 96413 tuning due to contamination from a strong adjacent transition of C$^{34}$S at 96416 MHz. Likewise we were not able to obtain upper limits for the 113156 transition of glycine I in Sgr B2 due to contamination in the 113153 tuning, due to C$^{13}$O contamination from the lower side band. However, in tunings 94055 (Orion KL and Sgr B2 (LMH)) and 113153 (Orion KL) we had clear portions of spectra at the expected positions of the transitions and so were able to obtain good upper limits (Table \ref{tab5}). We find upper limits for glycine I abundance of 3.7 $\times 10^{14}$ cm$^{-2}$ in both Orion-KL and Sgr B2 (LMH). This 3--sigma limit is of the same order as the reported detection of conformer I by \citet{k+2003} of 4.4 $\times 10^{14}$ cm$^{-2}$ in Orion--KL and 4.2 $\times 10^{14}$ cm$^{-2}$ in Sgr B2 (LMH). However, our results are 3--sigma upper limits rather than detections, and there is no evidence in the three good spectra for a weak feature at the expected positions. We conclude that this weighs against confirmation of the claimed detection of \citet{k+2003}. 

We find upper limits for glycine II abundance of 7.7 $\times 10^{12}$ cm$^{-2}$ in both Orion-KL and Sgr B2 (LMH). These are similar to the results of Combes, Rieu \& Wlodarczak (1996) who used the IRAM 30-metre telescope to search for glycine conformer II in a number of molecular clouds, including Sgr B2 and Orion A. The limits obtained by \citet{combes96} for glycine conformer II were 8$\times 10^{12}$ cm$^{-2}$ in Orion A and 1 $\times 10^{13}$ cm$^{-2}$ in Sgr B2. In both Orion KL and Sgr B2 (LMH), our upper limits are over an order of magnitude lower than the reported detections of  conformer I \citep{k+2003}, who used observations from the NRAO 12-m telescope, which like the Mopra Telescope will be most sensitive to extended emission. The results presented here show that glycine conformer II is not present in the extended gas at the level detected by \citet{k+2003} for conformer I, and the presence of both conformers in the compact hot LMH at the level reported by \citet{k+2003} has previously been ruled out (Paper I; Snyder et al. 2005). Although we cannot completely rule out the presence of glycine conformer I in the extended gas, this scenario also seems unlikely, and we conclude that it is unlikely that glycine has been detected in the ISM. A summary of these results is given in Table \ref{tab6}.

The difference in the upper limits for glycine II compared to those of glycine I deserves some comment. It results partly from better sensitivity in some of the tunings which include glycine II transitions, but the major effect is the higher dipole moment of conformer II c.f. conformer I, as spectral intensity is proportional to the square of the dipole moment (see e.g. Paper 1). If we use the reported detections of \citet{k+2003} for conformer I (similar to our upper limits) then we find that the ratio of glycine conformer II to conformer I is $\leq$ 0.02 in both Orion-KL and Sgr B2. Given the high energy barrier between the two conformers, which
prevents ready conversion of one to another, this ratio reflects the history of the glycine molecules rather than the current gas temperature.  It suggests that the
reactions by which glycine is produced favour conformer I, unless conformer I has an abundance far lower than the upper limits presented in this paper.    

We have also calculated the upper limits for the column density
of propylene oxide (Table \ref{tab5}).
We have assumed excitation temperature 
$T_{ex} = 200$ K
and Q(200) = 17811 from the approximate formula
$ Q(T) = (kT/h)^{3/2} (\pi/ABC)^{1/2} $ and the rotational constants
A,B,C from Creswell \& Schwendeman (1977). The energy levels and Einstein 
coefficients (Table \ref{tab5}) were calculated with a rigid asymmetric rotor model, 
using the spectroscopic constants from Creswell \& Schwendeman (1977)
and the dipole moment components from Swalen \& Herschbach (1957). We find upper limits for propylene oxide abundance of 3.0 $\times 10^{14}$ cm$^{-2}$ in Orion-KL and 6.7 $\times 10^{14}$ cm$^{-2}$ in Sgr B2 (LMH), assuming the propylene oxide emission is extended with respect to the Mopra beam. In \citetalias{paper1} we reported upper limits for propylene oxide emission of 8.9 $\times 10^{15}$ cm$^{-2}$ in Sgr B2 (LMH), assuming that the emission is extended with respect to the ATCA synthesized beam of 17.0 $\times$ 3.4 arcsec$^2$, so the Mopra observations place a more stringent limit on the presence of propylene oxide in the extended gas of Sgr B2. Ethylene oxide, the lower homologue of propylene oxide, was found by \citet{dickens97} to have an abundance of 3.3 $\times 10^{14}$ cm$^{-2}$ in Sgr B2(N), with the emission likely to be originating from extended gas rather than a compact source. As it is unlikely that propylene oxide will be more abundant than ethylene oxide, the non--detection at the level of sensitivity achieved in the Mopra
observations is to be expected.

\section{Unassigned Features}

We have detected fourteen features in Sgr B2 and four features in Orion-KL not previously reported in the ISM (Tables and Figures \ref{tab3} and \ref{tab4}). Some features are almost certainly the same carrier seen in the two separate velocity components of 64 and 73 km s$^{-1}$ present near Sgr B2 (LMH). UK86944 has a velocity separation of +10.9 km s$^{-1}$ from UK86947 and so these are likely to be same transition seen in both components, as is likely to be the case for UK86979 and UK86976 (separated by 10.9 km s$^{-1}$). This also may be the case for UK113128 and UK113126 (separation 5.3 km s$^{-1}$) although this spectrum is contaminated by $^{13}$CO in the lower side band so we have less confidence in the accuracy of the line centres. 

The features UK86979, UK86976 and UK86956 were also observed with the ATCA with a synthesized beam of 17.0 $\times$ 3.4 arcsec$^2$. These transitions are clearly seen in the ATCA spectrum, and have similar line profiles \citepalias{paper1}. For all three transitions the brightness temperature of the Mopra observations compared with those of the ATCA are consistent with the carriers of these features being found in the hot core of the LMH rather than in the extended gas. 

The previously undetected features were checked against line frequencies from the Jet Propulsion Laboratory (JPL, Pickett et al. 1998) and Cologne  Database for Molecular Spectroscopy (CDMS, M\"{u}ller et al 2001) on-line databases. Although a number of features were found to have coincident entries in the databases, all possibilities were ruled out by calculating the column density of the molecule that detection of the particular transition would imply, using the procedure described in section \ref{limits}.    

\section{Conclusions}

We have used the Mopra Telescope to search for glycine conformers I and II, and the simple chiral molecule propylene oxide, in Sgr B2 (LMH) and Orion KL. The major results are as follows:

\begin{itemize}

\item We did not detect either molecule but have been able to place sensitive upper limits on the column density of both.

\item We find upper limits for glycine I abundance of 3.7 $\times 10^{14}$ cm$^{-2}$ in both Orion-KL and Sgr B2 (LMH). In both sources this is comparable to the reported detections of  conformer I  by \citet{k+2003}.

\item We find upper limits for glycine II abundance of 7.7 $\times 10^{12}$ cm$^{-2}$ in  both Orion-KL and Sgr B2 (LMH). These are similar to the results of \citet{combes96}, and imply a ratio of glycine conformer II to conformer I of $\leq$ 0.02 in both Sgr B2 and Orion--KL, if the results of \citet{k+2003} were to be confirmed.

\item The results presented here show that glycine conformer II is not present in the extended gas at the levels detected by \citet{k+2003} for conformer I. While we cannot rule out the presence of glycine conformer I in the extended gas as our upper limits are very similar to the column density reported, our results are 3--$\sigma$ upper limits rather than detections, and we conclude that this weighs against confirmation of the detection of \citet{k+2003}. The detection of glycine in the compact hot core of the LMH at the levels reported has previously been ruled out, so we conclude that the evidence is against \citet{k+2003} having detected glycine in the ISM.

\item We find upper limits for propylene oxide abundance of 3.0 $\times 10^{14}$ cm$^{-2}$ in Orion-KL and 6.7 $\times 10^{14}$ cm$^{-2}$ in Sgr B2 (LMH). These limits are applicable assuming  that the propylene oxide emission is extended. In \citetalias{paper1} we reported upper limits for propylene oxide emission of 8.9 $\times 10^{15}$ cm$^{-2}$ in Sgr B2 (LMH), assuming that the emission is extended with respect to the ATCA synthesized beam of 17.0 $\times$ 3.4 arcsec$^2$, so the Mopra observations place a more stringent limit on the presence of propylene oxide in the extended gas of Sgr B2.    

\item We have detected fourteen features in Sgr B2 and four features in Orion-KL not previously reported in the ISM, but have not be able to plausibly assign these transitions to carriers.

\end{itemize}

\section*{Acknowledgments}

The Monash authors acknowledge the financial
support provided by the Australia Telescope National Facility for the construction of the laboratory spectrometer, the spectrometer design, construction and technical
services provided by Jonathan G. Crofts, and the collaboration with Takeshi  
Sakaizumi in the laboratory measurements of the Glycine spectrum. MRC and PAJ would like to thank the Max--Planck--Institut f\"{u}r Radioastronomie, Bonn, Germany for support and facilities provided during the preparation of the paper.
PAJ would like to acknowledge the support provided by  an Australia Telescope National Facility Visiting Fellowship.
%  We thank XXXXX for carrying out some of the Mopra observations.
% We thank XXXXX for some helpful suggestions,
% XXXX for a critical reading of the original version of the
% paper and an anonymous referee for very useful comments that improved
% the presentation of the paper.

%%%%%%%%%%%%%%%%%%%%%%%%%%%%%%%%%%%%%%%%%%%%%%%%%%%%
% \appendix

\bsp

\label{lastpage}

\clearpage
% flush out floats for draft document

\end{document}